\documentclass[10pt, conference, letterpaper]{IEEEtran}
\usepackage[T1]{fontenc}
\usepackage[utf8]{inputenc}
\usepackage{microtype}
\usepackage{pifont}

\usepackage{titlesec}
\usepackage{amsmath} 

\usepackage{hyperref}

\titlespacing*{\section}{0pt}{1ex}{0.6ex}
\titlespacing*{\subsection}{0pt}{0.8ex}{0.5ex}

\usepackage{amsmath,amssymb}
\DeclareMathOperator*{\argmax}{arg\,max}

\usepackage{blkarray}
\usepackage{bm} 

\usepackage{booktabs,multirow, array, enumitem}
\usepackage{tabularx}

\usepackage{graphicx}
\usepackage[caption=false,font=footnotesize,subrefformat=parens,labelformat=parens]{subfig}
\usepackage{capt-of}
\usepackage{xcolor}
\usepackage{siunitx}

\usepackage[most,breakable]{tcolorbox}

\usepackage{algorithm}
\usepackage{algpseudocode}

\usepackage{xurl}

\usepackage[noadjust]{cite}

\usepackage[authormarkup=superscript,deletedmarkup=sout,addedmarkup=em,commandnameprefix=always]{changes}
\usepackage{soul}
\colorlet{soulcyan}{cyan!30}
\colorlet{soulgreen}{green!30}

\definechangesauthor[name={Claudio Fiandrino}, color=green!75!black]{CF}
\definechangesauthor[name={Joerg Widmer}, color=blue]{JW}
\soulregister\cite7
\soulregister\ref7
\soulregister\pageref7

\usepackage{xspace}

\usepackage{enumitem}
\newcommand{\simpletitle}[1]{\noindent\textbf{#1}.\xspace}

\DeclareTextFontCommand{\mytexttt}{\ttfamily\hyphenchar\font=45\relax}

\newcommand{\toolname}{\textsc{SymbXRL}\xspace}

\newcommand{\explora}{\textit{EXPLORA}\xspace}

\newcommand\txbrate{\relax\ifmmode\mathtt{tx\_brate}\else\texttt{tx\_b\-rate}\fi\xspace}
\newcommand\txpkts{\relax\ifmmode\mathtt{tx\_pkts}\else\texttt{tx\_pkts}\fi\xspace}
\newcommand\dlbuff{\relax\ifmmode\mathtt{dl\_buffer}\else\texttt{dl\_buffer}\fi\xspace}

\usepackage{xurl}
\usepackage{tikz} 
\usepackage[printonlyused]{acronym}

\acrodef{ai}[AI]{Artificial Intelligence}
\acrodef{ml}[ML]{Machine Learning}
\acrodef{dl}[DL]{Deep Learning}
\acrodef{aml}[AML]{Adversarial Machine Learning}
\acrodef{xai}[XAI]{EXplainable Artificial Intelligence}
\acrodef{nn}[NN]{Neural Networks}
\acrodef{dnn}[DNN]{Deep Neural Networks}
\acrodef{rnn}[RNN]{Recurrent Neural Networks}
\acrodef{lstm}[LSTM]{Long-Short Term Memory}
\acrodef{gnn}[GNN]{Graph Neural Networks}
\acrodef{shap}[SHAP]{SHapely Additive exPlanations}
\acrodef{lime}[LIME]{Local Interpretable Model-agnostic Explanations}
\acrodef{xgb}[XGBoost]{Extreme Gradient Boosting}
\acrodef{lrp}[LRP]{LayeR-wise backPropagation}
\acrodef{nlp}[NLP]{Natural Language Processing}
\acrodef{tcn}[TCN]{temporal convolutional network}
\acrodef{gpr}[GPR]{Gaussian process regression}
\acrodef{dt}[DT]{Decision Tree}
\acrodefplural{DT}[DTs]{Decision Trees}
\acrodef{xrl}[XRL]{EXplainable Reinforcement Learning}
\acrodef{rl}[RL]{Reinforcement Learning}
\acrodef{drl}[DRL]{Deep Reinforcement Learning}
\acrodef{crl}[CRL]{Casual Reinforcement Learning}
\acrodef{dqn}[DQN]{Deep Q-Network}
\acrodef{ppo}[PPO]{Proximal Policy Optimization}
\acrodef{a3c}[A3C]{Asynchronous Advantage Actor-Critic}
\acrodef{fol}[FOL]{First-Order Logic}
\acrodef{mdp}[MDP]{Markov Decision Process}
\acrodef{sac}[SAC]{Soft Actor-Critic}

\acrodef{fgsm}[FGSM]{Fast Gradient Sign Method}
\acrodef{bim}[BIM]{Basic Iterative Method}
\acrodef{mae}[MAE]{Mean Absolute Error}
\acrodef{rmse}[RMSE]{Root Mean Square Error}

\acrodef{qos}[QoS]{Quality of Service}
\acrodef{sla}[SLA]{Service Level Agreement}
\acrodef{isp}[ISP]{Internet Service Providers}
\acrodef{bs}[BS]{Base Station}
\acrodefplural{BS}[BSs]{Base Stations}
\acrodef{qos}[QoS]{Quality of Service}
\acrodef{gnb}[gNB]{next Generation Node B}
\acrodefplural{gNB}[gNBs]{next Generation Node Bs}
\acrodef{enb}[eNB]{evolved Node B}
\acrodefplural{eNB}[eNBs]{evolved Node Bs}
\acrodef{ue}[UE]{User Equipment}
\acrodef{prb}[PRB]{Physical Resource Block}
\acrodefplural{PRB}[PRBs]{Physical Resource Blocks}
\acrodef{mcs}[MCS]{Modulation and Coding Scheme}
\acrodef{tti}[TTI]{Transmission Time Interval}
\acrodef{rnti}[RNTI]{Radio Network Temporary Identifier}
\acrodefplural{RNTI}[RNTIs]{Radio Network Temporary Identifiers}
\acrodef{tbs}[TBS]{Tranport Block Size}
\acrodef{rrc}[RRC]{Radio Resource Control}
\acrodef{ran}[RAN]{Radio Access Network}
\acrodef{lmf}[LMF]{Location Management Function}
\acrodef{amf}[AMF]{Access and Mobility Function}
\acrodef{vnf}[VNF]{Virtual Network Function}
\acrodef{embb}[eMBB]{enhanced Mobile BroadBand}
\acrodef{mmtc}[mMTC]{massive Machine-type Communications}
\acrodef{urllc}[URLLC]{Ultra-Reliable and Low Latency Communications}

\acrodef{ias}[IAS]{Intent-based Action Steering}
\acrodef{as}[AS]{Action-Steering}
\acrodef{kpi}[KPI]{Key Performance Indicator}
\acrodef{kg}[KG]{Knowledge Graph}
\acrodefplural{KPI}[KPIs]{Key Performance Indicators}

\acrodef{llm}[LLM]{Large Language Model}
\acrodefplural{LLM}[LLMs]{Large Language Models}
\acrodef{chkp}[CHKP]{Checkpoint}
\acrodefplural{CHKP}[CHKPs]{Checkpoints}

\acrodef{dtu}[DTU]{Data Transmitted of User}
\acrodef{mse}[MSE]{Maximum Available Spectral Efficiency}
\acrodef{g}[G]{User Group Label}
\acrodef{los}[LoS]{Line of Sight}
\acrodef{nlos}[NLoS]{Non-Line of Sight}

\begin{document}

\bstctlcite{IEEEexample:BSTcontrol}
\title{\toolname: Symbolic Explainable Deep Reinforcement Learning for Mobile Networks \vspace*{-10pt}}

\author{Abhishek Duttagupta$^{*\dagger\Diamond}$, MohammadErfan Jabbari$^{*\Diamond}$, Claudio Fiandrino$^{*}$, Marco Fiore$^{*}$ and Joerg Widmer$^{*}$\\
$^{*}$IMDEA Networks Institute, Spain, $^{\dagger}$Universidad Carlos III de Madrid, Spain\\
Email: \{name.surname\}@imdea.org\vspace*{-15pt}
}

\setlength{\columnsep}{0.19in}
\maketitle
\begingroup
\renewcommand\thefootnote{$\Diamond$}
\footnotetext{These authors contributed equally to this work.}
\endgroup

\newcommand\publishedtext{%
	\footnotesize This is the author's accepted version of the article. The final version published by IEEE is A. Duttagupta, M. Jabbari, C. Fiandrino, M. Fiore, and J. Widmer, ``\toolname: Symbolic Explainable Deep Reinforcement Learning for Mobile Networks," IEEE INFOCOM 2025 - IEEE Conference on Computer Communications, doi: TBD.} 
\newcommand\copyrighttext{%
  \footnotesize \textcopyright 2025 IEEE. Personal use of this material is permitted.
  Permission from IEEE must be obtained for all other uses, in any current or future
  media, including reprinting/republishing this material for advertising or promotional
  purposes, creating new collective works, for resale or redistribution to servers or
  lists, or reuse of any copyrighted component of this work in other works.}
\newcommand\copyrightnotice{%
\begin{tikzpicture}[remember picture,overlay]
\node[anchor=north,yshift=0pt] at (current page.north) {\fbox{\parbox{\dimexpr\textwidth-\fboxsep-\fboxrule\relax}{\publishedtext}}};
\node[anchor=south,yshift=10pt] at (current page.south) {\fbox{\parbox{\dimexpr\textwidth-\fboxsep-\fboxrule\relax}{\copyrighttext}}};
\end{tikzpicture}}
\copyrightnotice

\begin{abstract}
The operation of future 6th-generation (6G) mobile networks will increasingly rely on the ability of~\ac{drl} to optimize network decisions in real-time.~\ac{drl} yields demonstrated efficacy in various resource allocation problems, such as joint decisions on user scheduling and antenna allocation or simultaneous control of computing resources and modulation. However, trained~\ac{drl} agents are closed-boxes and inherently difficult to explain, which hinders their adoption in production settings.
In this paper, we make a step towards removing this critical barrier by presenting \toolname, a novel technique for~\ac{xrl} that synthesizes human-interpretable explanations for~\ac{drl} agents. \toolname leverages symbolic AI to produce explanations where key concepts and their relationships are described via intuitive symbols and rules; coupling such a representation with logical reasoning exposes the decision process of~\ac{drl} agents and offers more comprehensible descriptions of their behaviors compared to existing approaches. We validate \toolname in practical network management use cases supported by~\ac{drl}, proving that it not only improves the semantics of the explanations but also paves the way for explicit agent control: for instance, it enables intent-based programmatic action steering that improves by 12\% the median cumulative reward over a pure~\ac{drl} solution.
\end{abstract}

\section{Introduction}%
\label{sec:intro}%

The future 6th generation (6G) of mobile networks promises unprecedented connectivity with ultra-high data rates, minimal latency, and massive device support \cite{saad2019vision}. Recent reports predict global mobile network data traffic will reach 466 exabytes per month by 2029, driven by the proliferation of connected devices and data-intensive applications \cite{ericsson-report-2024}. As network complexity increases,~\ac{ai} emerges as a crucial enabler for network management and optimization~\cite{letaief2019roadmap}. The integration of~\ac{ai} is expected to enhance network efficiency and enable new services and applications~\cite{wikstrom2020challenges}. 

A specific~\ac{ai} paradigm that shows significant promise for 6G is~\ac{drl}, which combines~\ac{dl} with~\ac{rl} to tackle complex mobile network challenges~\cite{luong2019applications}.~\ac{drl} enables agents to learn optimal policies through interaction with their environment and has been successfully applied to resource allocation~\cite{wei2018joint} and network slicing~\cite{bega2020ai} in 5G systems. More advanced~\ac{drl} applications are emerging for 6G, including dynamic spectrum management~\cite{naparstek2018deep} and multi-agent coordination for network slicing~\cite{kim2021multi}, showcasing potential throughout the network ecosystem~\cite{sun2020machine}.

However, these~\ac{drl} applications operate as closed boxes that inherently lack interpretability, which makes debugging and troubleshooting hard~\cite{morovati2024common}, may compromise performance~\cite{heuillet2021explainability} and generally curbs adoption by network providers~\cite{auric}. Recent advances in~\ac{xai} address this lack of transparency
, yet current~\ac{xai} solutions for~\ac{drl} agents, such as \explora~\cite{explora}, METIS~\cite{metis-meng2020interpreting} and post-hoc interpretation methods like SHAP~\cite{shap-lundberg2017unified} and LIME~\cite{lime-kdd} often fall short in providing meaningful, human-understandable explanations for complex network management systems~\cite{milani2024explainable}, as we will also detail in our study.

In this paper, we propose \toolname, a novel~\ac{xrl} technique leveraging symbolic AI, specifically~\ac{fol}, to synthesize comprehensible explanations with rich semantics from symbolic representations of the states and actions of~\ac{drl} agents. \toolname employs~\ac{fol} to formalize the agent's behavior and decision-making process, allowing for more intuitive and interpretable explanations. This approach serves three primary objectives: 
\textit{(1)} providing simple, logically structured explanations to understand and compare~\ac{drl} agents; 
\textit{(2)} enabling~\ac{ias} through logical rules in the form of~\ac{fol} representation; and 
\textit{(3)} leveraging symbolic knowledge to identify flaws in the design process of new agents by analyzing logical inconsistencies or unexpected patterns in the agent's behavior. By translating complex numerical states and actions into symbolic logic statements, \toolname offers a unique perspective on~\ac{drl} agent behavior, bridging the gap between low-level operations and high-level, human-interpretable concepts and addressing current limitations of~\ac{xrl} solutions.

We validate our approach with two distinct~\ac{drl} use cases addressing critical 5G and 6G challenges. In the first application, a~\ac{drl} agent~\cite{polese2022colo} controls~\ac{ran} slicing and scheduling on a~\ac{gnb} as an O-RAN compliant xApp for three network slices serving different traffic categories. The second use case employs a~\ac{drl} agent~\cite{an2023deep} for resource scheduling in Massive MIMO. These agents offer diverse decision-making contexts: the first has a multi-modal action space with both continuous and discrete factors, where the actions affect all of its next state input~\ac{kpi}s; the second agent has a discrete action space that affects a subset of the input~\acp{kpi}. This diversity allows showcasing the flexibility of \toolname across different scenarios.

The key contributions (``C'') and findings (``F'') of our study are summarized as follows.
{%
\setlist[itemize]{leftmargin=1.75em}
\begin{itemize}

    \item[C1.] We propose \toolname, a novel explainer for~\ac{drl} agents using symbolic representations with~\ac{fol} to synthesize human-interpretable explanations.

    \item[C2.] We validate \toolname in two diverse use cases that rely on~\ac{drl} agents for network slicing and Massive MIMO scheduling, and demonstrate the superior intelligibility and detailed level of our solution compared to state-of-the-art approaches.
    
    \item[C3.] For reproducibility and to further stimulate the research in the field, we release the artifacts of our study (code of \ac{rl} agents and \toolname) at:https://github.com/RAINet-Lab/symbxrl.

    \item[F1.] We prove that \toolname provides human-readable and comprehensible symbolic explanations, which improve explanations of state-of-the-art methods and bridge the gap between DRL agent behavior and human understanding.

    \item[F2.] We show that \toolname's symbolic representation enables flexible \ac{ias} policies, which ($i$) improve the cumulative reward of~\ac{drl} agents and ($ii$) enable enforcement of operational constraints on the agent actions. Our experiments demonstrate that these capabilities result in a 12\% median improvement in cumulative reward over baseline performance, which outperforms existing~\ac{xrl} methods like METIS.

\end{itemize}
}

\section{Background and Technical Foundations}%
\label{sec:back-and-tech}%

This section provides the necessary background and technical foundations for understanding our proposed approach. We cover the key concepts of~\ac{rl} and~\ac{drl}, explainability in~\ac{ai}, and Symbolic \ac{ai} with a focus on~\ac{fol}.

 \subsection{Reinforcement Learning and Deep Reinforcement Learning}
\label{subsec:rl-drl}

Reinforcement Learning (\ac{rl}) is a computational approach to learn from interaction with an environment. \ac{rl} agents learn optimal policies by taking actions in an environment to maximize cumulative reward signal. The \ac{rl} process is typically modeled as a Markov Decision Process (MDP), defined by the tuple $(S, A, P, R, \gamma)$, where $S$ is the set of states, $A$ is the set of actions, $P(s_{t+1} | s_t, a_t)$ are state transition probabilities, $R(s_t, a_t)$ is the reward, and $\gamma \in [0, 1]$ is the discount factor.

At each time step $t$, the agent observes the current state $s_t \in S$, takes an action $a_t \in A$ following a policy $\pi : S \to P(A)$, receives a reward $r_t$, and transitions to the next state $s_{t+1}$. The agent's goal is to learn an optimal policy $\pi^*$ that maximizes the expected cumulative discounted reward:

\begin{equation}
\pi^* = \operatorname*{arg\,max}_\pi \mathbb{E}\left[\sum_{t=0}^{\infty} \gamma^t r_t | \pi\right].
\end{equation}

\ac{drl} extends \ac{rl} by utilizing \ac{dnn} to approximate value functions or policies, enabling \ac{rl} to scale to high-dimensional state and action spaces~\cite{mnih2015human}. \ac{drl} has been successfully applied to various domains, including playing complex games~\cite{vinyals2019grandmaster} and robotic control~\cite{haarnoja2023learning}.

\subsection{Explainability and XAI}
\label{subsec:xai}

\ac{xai} refers to techniques and methods that make the behavior of \ac{ai} systems comprehensible to humans. The primary goal of \ac{xai} is to create models that can provide clear and understandable explanations for their decisions, facilitating trust and adoption in critical applications~\cite{gunning2019darpa}.

In the context of \ac{drl}, explainability is particularly challenging due to the complex interactions between the agent and the environment over time~\cite{heuillet2021explainability}. To address these challenges, Explainable Reinforcement Learning (\ac{xrl}) techniques have been developed to provide insights into the decision-making process of \ac{drl} agents~\cite{puiatta-xrl-survey}. \ac{xrl} methods can be either \textit{intrinsic} when they modify the \ac{rl} algorithm itself to be explainable~\cite{juozapaitis2019explainable} or \textit{post-hoc} if the explanations are produced without altering the original model~\cite{shap-lundberg2017unified,lime-kdd,metis-meng2020interpreting,explora}.

Existing \ac{xrl} methods, however, often fall short in providing meaningful, human-understandable explanations for complex network management systems~\cite{dazeley-xrl}. METIS \cite{metis-meng2020interpreting}, an explainer using a mixture of \ac{dt} and hypergraphs, requires multiple rollouts of the whole \ac{rl} environment to improve the \ac{dt} efficiency, with a time complexity that makes this approach unsuitable for continuous, complex, multi-modal environments. \explora~\cite{explora}, an attribute graph-based explainer, uses low-level agent numerical data that complicates the synthesis of compact and easy-to-understand insights.

To address the shortcomings of the state of the art, we leverage \ac{fol}, a form of symbolic \ac{ai}, to explain the behavior and decision-making process of \ac{drl} agents.

\subsection{Symbolic AI and First-Order Logic}
\label{subsec:symbolic-ai}

Symbolic AI, in contrast to Statistical AI approaches like \ac{dl}, uses human-readable symbols and rules to represent knowledge and reason about it. First-Order Logic (\ac{fol}) is a powerful formalism within Symbolic AI that allows for the representation of complex relationships and reasoning.

\ac{fol} consists of the following key components:
\begin{tcolorbox}[colback=gray!15,colframe=gray!15,breakable,]
$\bullet$ \textbf{Constants:} Represent specific objects in the domain.\par
$\bullet$ \textbf{Variables:} Stand for arbitrary objects.\par
$\bullet$ \textbf{Predicates:} Express properties of objects or relationships between objects.\par
$\bullet$ \textbf{Quantifiers:} "For all" ($\forall$) and ``There exists'' ($\exists$).\par
$\bullet$ \textbf{Logical connectives:} AND ($\wedge$), OR ($\vee$), NOT ($\neg$), IMPLIES ($\Rightarrow$).
\end{tcolorbox}

To illustrate how \ac{fol} can represent policies in a networking context, consider the example: ``If a user's data consumption exceeds the plan's limit at 10 GB, throttle their connection speed.'' We can formalize this policy in \ac{fol} as follows.

\begin{itemize}
    \item \textbf{Predicates:}
    \begin{itemize}
        \item \( \text{Exceeds}(x, y) \): ``user \( x \)'s data usage exceeds \( y \) GB.''
        \item \( \text{Throttle}(x) \): ``throttle user \( x \)'s connection speed.''
    \end{itemize}
    \item \textbf{Constants and Variables:}
    \begin{itemize} 
        \item \( u \): a user.
        \item \( L \): the plan's limit, which is a constant value of 10 GB.
    \end{itemize}
    \item \textbf{\ac{fol} Statement:} $\forall u \, (\text{Exceeds}(u, L) \Rightarrow \text{Throttle}(u))$.
\end{itemize}

This \ac{fol} statement reads as: ``For all users \( u \), if user \( u \)'s data usage exceeds the plan's limit \( L \) (which the operator has set at 10 GB), then throttle user \( u \)'s connection speed.''

We chose \ac{fol} for our work due to its balance between expressiveness and simplicity. Unlike propositional logic, \ac{fol} can define predicates, variables, and constants, capturing the complexity of our agent's behavior. At the same time, it avoids the advanced features of higher-order logic that are beyond our current requirements~\cite{halpern2008using}.

By leveraging \ac{fol}, we can create a symbolic representation of \ac{drl} agents' behavior, enabling more intuitive and interpretable explanations of their decision-making processes~\cite{ma2021learning}. This approach bridges the gap between the low-level numerical operations of \ac{drl} and high-level, human-understandable concepts, addressing limitations of current \ac{xrl} techniques.

\section{\toolname}%
\label{sec:design}%

In this section, we describe the details of \toolname and provide a comprehensive overview of its framework. This includes creating symbolic representations for the state and action space of the agent, generating explanations, and integrating these representations into the explainability pipeline. We also describe how we leverage this symbolic representation to provide \ac{ias} for the agents.

\subsection{Overview of \toolname}
\label{subsec:overview}

\toolname leverages \ac{fol} to produce explanations for \ac{drl} agents. The symbolic representation offers the following three key advantages:

\begin{enumerate}[label=$\bullet$, wide=0\parindent, listparindent=0pt, align=left]
    \item \textit{Enhanced Interpretability:} By representing agent behavior in logical terms, \toolname enables formal reasoning techniques for analyzing and verifying agent behavior. This approach produces explanations that are more intuitive compared to previous works~\cite{metis-meng2020interpreting}.
    
    \item \textit{Concise Representation:} The symbolic representation provides a more concise and compact representation of the state-action space, facilitating easier analysis and visualization compared to state-of-the-art explainability tools~\cite{explora}.

    \item \textit{Flexible Action Steering:} This symbolic framework allows the definition of high-level rules to guide agent behavior, enhancing performance tuning and ensuring better compliance with operational constraints.
\end{enumerate}

\begin{figure}[t]
\centering%
\includegraphics[width=0.8\columnwidth,keepaspectratio]{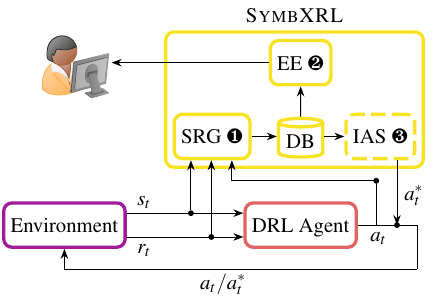}%
\vspace*{-1ex}%
\caption{\toolname's architecture and interaction with a \ac{drl} agent operating on a target environment.}%
\label{fig:xrlogicLens-architecture}%
\vspace*{-1ex}%
\end{figure}

Fig.~\ref{fig:xrlogicLens-architecture} illustrates \toolname's architecture and integration with \ac{drl} pipelines. The key components are:

\begin{enumerate}[label=\arabic*), wide=0\parindent, listparindent=0pt, align=left]
    \item \textbf{Symbolic Representation Generator (SRG~\ding{182}):} Transforms numerical states and actions into \ac{fol} terms and stores this information in a database.
    \item \textbf{Explanation Engine (EE~\ding{183}):} Utilizes symbolic representations to generate human-readable explanations of agent behavior in the form of a~\ac{kg}.
    \item \textbf{Intent-based Action Steering (IAS~\ding{184}):} This optional module enables real-time fine-tuning of agent behavior by translating high-level intents via \ac{fol} representation.
\end{enumerate}

\subsection{Symbolic Representation Generator~(\ding{182})}
\label{subsec:symb-repr}

The cornerstone of \toolname is creating symbolic representations for the \ac{drl} agent's state and action spaces, mapping numerical or categorical values into logical terms in \ac{fol}. The process begins by defining the explanation goal that guides the selection of relevant variables and the level of abstraction.

For each selected variable, we define \ac{fol} terms that capture essential information:
\begin{itemize}
    \item For continuous variables: \textbf{predicate}(variable, quartile), where the predicate indicates the direction of change (increase, decrease, or constant).
    \item For categorical variables: \textbf{toCategory}(variable).
\end{itemize}

The quartiles (Q1, Q2, Q3, Q4) represent the relative magnitude of the value, calculated efficiently using the P2 algorithm~\cite{p2-algo} for online computation of quartile markers. These symbolic representations are stored in a database (DB) for use by other components and services.

\simpletitle{An Example} To clarify, consider a \ac{drl} agent performing resource allocation on a \ac{gnb}. Its state might include the number of connected users (integer), average user throughput (float, Mbps), and interference level (categorical: Low, Medium, High). The action space could involve adjusting transmission power (float, dBm) and changing frequency band (categorical: Low, Mid, High). In this context, \textbf{inc}(users, Q4) represents a significant increase in user numbers, while \textbf{toHigh}(interf) indicates high interference levels. The choice of predicates and granularity balances expressiveness and complexity, optimized for the specific application and explanatory goal.

\subsection{Explanation Engine~(\ding{183})}
\label{subsec:gen-explanations}

The Explanation Engine (EE) leverages the symbolic representations created by the SRG module~(\ding{182}) to generate insights into the agent's behavior. This approach enables two analytical methods: \textit{probabilistic analysis} and \textit{\ac{kg} analysis}.

In \textit{probabilistic analysis}, the EE performs four main steps: collects symbolic representations of states and actions from the DB, counts occurrences of each unique symbolic state and action, calculates probabilities or frequencies of these occurrences, and visualizes the results through probability distributions, correlation density maps, or~\ac{kg}s. This provides insights into both input state distributions and the correlation of agent's actions and their effects effects on the environment.

In \textit{\ac{kg} analysis}, EE constructs a graph where nodes represent symbolic actions and edges are transitions between them. The weight of each node and edge corresponds to the frequency of that action and transition. \ac{kg} reveals the agent's learned decision-making strategies and overall behavior patterns.

\simpletitle{An Example} Applying these analyses to our toy example:
The correlation density map reveals that the agent maintains average throughput in Q3 (\textbf{const}(throughput, Q3)) by keeping transmission power in Q4 (\textbf{const}(tx\_power, Q4)). The~\ac{kg} shows the agent frequently switches between mid and high frequency bands, represented as transitions between \textbf{toMid}(freq\_band) and \textbf{toHigh}(freq\_band) nodes.

These analytical approaches offer several advantages over state-of-the-art methods~\cite{explora,metis-meng2020interpreting}: ($i$) direct insights into the agent's decision-making process, ($ii$) revealing patterns not apparent from reward analysis, and ($iii$) enabling comparison between different agents or versions. While our framework generalizes to various \ac{drl} agents and environments, the choice of symbolic representations and \ac{fol} definitions is use-case specific. Though discretization may lead to some loss of detail, a well-defined explanation goal can mitigate this limitation.

By combining these analytical approaches with symbolic representation, the EE~\ding{183} module generates intelligible explanations of the agent's behavior, bridging the gap between closed-box~\ac{drl} agents and human understanding.

\subsection{Intent-based Action Steering~(\ding{184})}
\label{subsec:intent-action-steering}

We introduce Intent-based Action Steering (\ac{ias}), a mechanism integrating symbolic representation to guide agent behavior towards network operators' specific intents. Inspired by decision shielding in reinforcement learning~\cite{rl-shielding-alshiekh-aaai-2018}, \toolname's \ac{ias} leverages symbolic knowledge to surpass traditional methods. Unlike approaches that may reduce rewards or introduce unfamiliar states or actions, \ac{ias} selects actions from the agent's prior experiences, ensuring both constraint satisfaction and performance optimization.

\toolname's unique \ac{ias} approach operates on discretized and concise state and action spaces, unlike previous methods such as \explora~\cite{explora}. This design offers two key advantages: it \textit{(i)} enables efficient and accurate matching between current and past states and actions, reducing computational complexity; \textit{(ii)} allows operators to define intents using the same \ac{fol} format as agent explanations, seamlessly integrating intents without compromising learned behavior.

We demonstrate the versatility of \toolname's \ac{ias} through three distinct use cases:

\begin{enumerate}[wide=0\parindent, listparindent=0pt, align=left]
    \item \textit{Reward maximization:} enhances the cumulative reward of the agent by maximizing each step's reward through targeted action steering ($a^*_t$), achieving:
        \begin{equation}
            \label{eq:reward-maximizing-formula}
            a^*_{t} = \argmax_{a_1, \ldots, a_T} \sum_{t=1}^T r_t(s_t, a_t) \text{ s.t. } a_t \in \mathcal{A}(s_t).
        \end{equation}
        where $\mathcal{A}(s_t)$ is the set of actions the agent has previously taken for state $s_t$ plus the current timestep action. $a^*_t$ is the optimal action chosen to maximize the reward at each step.

    \item \textit{Decision conditioning:} applies constraints to the agent's actions to enforce operational limits or policy requirements without excluding reward maximization. For example:\vspace*{-.5ex}
        \begin{tcolorbox}[colback=gray!15,colframe=gray!15,breakable,boxsep=.1mm, left=3pt]
            $\bullet$ Schedule a specific user: $\textbf{Schedule}(a_t, \text{UserID})$
            
            $\bullet$ Do not schedule users in a group: $\forall~\text{UserID} \in \text{Group} : \textbf{notSchedule}(a_t, \text{UserID})$
        \end{tcolorbox}
    These examples illustrate the flexibility in defining policies beyond simple reward or throughput maximization~\cite{explora}. This mode allows us to improve the agent's cumulative reward while applying operational constraints.

    \item \textit{Accelerated learning:} Reducing training time for \ac{drl} agents is crucial for faster deployment and resource efficiency~\cite{acc-learning-sample-efficiency-neurips-2021}. By effectively leveraging the agent's acquired knowledge, \ac{ias} allows agents trained for fewer episodes to achieve competitive cumulative rewards compared to those trained for longer periods.
\end{enumerate}

\simpletitle{An Example} Applying \ac{ias} for reward maximization to our toy example:
Given the environment's input state and agent's action, \ac{ias} checks the~\ac{kg} for a potentially better action (transmission power and frequency band) that could yield a higher expected reward. This replacement action is chosen based on the similarity between the current state and the state where the action was previously applied.

\begin{table*}
    \setlength{\columnsep}{0.3in}  
    \centering%
    \caption{State and action spaces and their symbolic representations of the two~\ac{drl} agents used for validation}%
    \label{tab:agent_info_comparison}%
    \vspace*{-1ex}%
    \begin{tabular}{p{3.2cm}p{6.9cm}p{6.9cm}}  
        \toprule
        Agents & 
        \textbf{A1: Network Slicing and Scheduling Agent} & 
        \textbf{A2: Massive MIMO Scheduling Agent} \\
        \midrule
        State Space & 
        $s_t \in \mathcal{S} = \mathbb{R}^{M \times K \times |\mathcal{L}|}$ where:
        \begin{itemize}[leftmargin=*, topsep=0pt, itemsep=0pt]
            \item $\mathcal{M}$: Measurements, $M = 10$
            \item $\mathcal{L}$: Slices, $\mathcal{L} = \{\text{\ac{embb}, \ac{mmtc}, \ac{urllc}}\}$
            \item $\mathcal{K}$: \acp{kpi}, $\mathcal{K} = \{\text{tx\_brate, tx\_pkts, dl\_buffer}\}$
        \end{itemize} & 
        $s_t \in \mathcal{S} = \mathbb{R}^{K \times |\mathcal{N}|}$ where:
        \begin{itemize}[leftmargin=*, topsep=0pt, itemsep=0pt]
            \item $\mathcal{N}$: Number of Users, $N = 7$
            \item $\mathcal{K}$: \acp{kpi}, $\mathcal{K} = \{\text{MSE, DTU, G}\}$
        \end{itemize} 
        \\[-2.5ex]
        \midrule
        Symbolic Representation of State Space & 
        For each $k \in \mathcal{K}$ and $l \in \mathcal{L}$:
        \begin{itemize}[leftmargin=*, topsep=0pt, itemsep=0pt]
            \item $\bar{k}_l \rightarrow \text{Pred}(\bar{k}_l, Q)$, where $\bar{k}_l = \frac{1}{M}\sum_{m=1}^M k_{l,m}$
            \item Pred $\in$ Predicate $= \{\text{inc, dec, const}\}$
            \item Q $\in$ Quartile $= \{\text{Q1, Q2, Q3, Q4}\}$
            \item $M = 10$: Number of measurements
        \end{itemize} & 
        For each $k \in \mathcal{K}$ and $g \in \mathcal{G}$:
        \begin{itemize}[leftmargin=*, topsep=0pt, itemsep=0pt]
            \item $\bar{k}_g \rightarrow \text{Pred}(\bar{k}_g, Q)$, where $\bar{k}_g = \frac{1}{|U_g|}\sum_{u \in U_g} k_u$
            \item Pred $\in$ Predicate $= \{\text{inc, dec, const}\}$
            \item Q $\in$ Quartile $= \{\text{Q1, Q2, Q3, Q4, MAX}\}$
            \item $U_g$: Set of users in group $g$
        \end{itemize}
        \\[-2.5ex]
        \midrule
        Action Space & 
        $a_t \in \mathcal{A} = \text{\ac{prb}}^{|\mathcal{L}|} \times \text{SP}^{|\mathcal{L}|}$ where:
        \begin{itemize}[leftmargin=*, topsep=0pt, itemsep=0pt]
            \item \ac{prb}: Physical Resource Block $= \{1, 2, \ldots, 50\}$
            \item SP: Scheduling Policy $= \{\text{WF, RR, PF}\}$
            \item $\mathcal{L}$: Slices
        \end{itemize} & 
        $a_t \in \mathcal{A} = \{0,1\}^N$ where:
        \begin{itemize}[leftmargin=*, topsep=0pt, itemsep=0pt]
            \item $N$: Number of users, $N = 7$
            \item $0$: Do not schedule a user
            \item $1$: Schedule a user
        \end{itemize} \\[-2.5ex]
        \midrule
        Symbolic Representation of Action Space & 
        For each $l \in \mathcal{L}$:
        \begin{itemize}[leftmargin=*, topsep=0pt, itemsep=0pt]
            \item $\text{\ac{prb}}_l \rightarrow \text{Pred}(\text{\ac{prb}}, \text{C}_{\text{start}}, \text{C}_{\text{final}})$
            \item $\text{SP}_l \rightarrow \text{toPolicy}(\text{sched})$
            \item Pred: Predicate $\in \{\text{inc, dec, const}\}$
            \item C: \ac{prb} Category $\in \{\text{C1, C2, C3, C4, C5}\}$
            \item toPolicy $\in \text{Scheduling Policy} = \{\text{toWF, toRR, toPF}\}$
        \end{itemize} & 
        For each $g \in \mathcal{G}$:
        \begin{itemize}[leftmargin=*, topsep=0pt, itemsep=0pt]
            \item $a_g \rightarrow \text{Pred}(g, Q, Percentage)$
            \item Pred: Predicate $\in \{\text{sched}\}$
            \item $g$: Group number, $g \in \mathcal{G}$
            \item Q: Quartile $\in \{\text{Q1, Q2, Q3, Q4, MAX}\}$
            \item Percentage: $\text{Round}_{\{0, 25,..., 100\}}\left(\frac{|\text{Sched. users in } g|}{|\text{Total users in } g|} \times 100\right)$
        \end{itemize} \\[-2.5ex]
        \bottomrule
    \end{tabular}
    \vspace*{-3ex}%
\end{table*}

\section{Application to Practical Use Cases}%
\label{sec:evaluation}%

The operation and advantages of \toolname are best demonstrated through practical applications. This section presents a comprehensive empirical evaluation of \toolname using two~\ac{rl} use cases for network management. We detail the evaluation framework and experimental setup~(\S\ref{subsec:framework} and~\S\ref{subsec:setup}), followed by our findings on~\ac{drl} agent behavior~(\S\ref{subsec:insights}) and performance improvements achieved through~\ac{ias}~(\S\ref{subsec:action-steering}).

\subsection{Evaluation Framework}
\label{subsec:framework}

Our evaluation of \toolname focuses on two main objectives. First, we assess the explanation quality and interpretability through a qualitative analysis of compactness and clarity of the generated explanations. Second, we quantify performance improvements achieved with \ac{ias} by measuring changes in cumulative reward and training efficiency.

\subsection{Symbolic Representations for the Agents}
\label{subsec:setup}

To validate \toolname, we employ two distinct~\ac{drl} agents addressing challenging resource allocation and scheduling problems in mobile networks. By selecting these agents, we demonstrate the versatility of \toolname. \textit{Table~\ref{tab:agent_info_comparison}} summarizes state-action space and the~\ac{fol} representations for each agent.

\begin{enumerate}[label=\textbf{A\arabic*}, wide=0\parindent, listparindent=0pt, align=left]

\item \label{agent-neu} \textbf{Network Slicing and Scheduling Agent} \cite{polese2022colo}: 
    This agent jointly controls \ac{ran} slicing and scheduling policies for three slices $\mathcal{L} = \{\text{\ac{embb}, \ac{mmtc}, \ac{urllc}}\}$ in an OpenAI Gym environment with O-RAN compliant xApps. The agent is trained and evaluated in the Colosseum emulator using two traffic profiles: TRF1 (slice-based) and TRF2 (uniform). It operates in two modes, each favoring one slice (i.e., \ac{embb} or \ac{urllc}) by adjusting the weight of each slice's \ac{kpi} effect in the reward function (\txbrate for~\ac{embb}~and~\dlbuff for~\ac{urllc}).

    For \ac{fol}-based symbolic representation, we use the format \textbf{predicate}(variable, Quartile) for continuous, unbounded variables (e.g., \acp{kpi}). For bounded discrete variables (e.g., \ac{prb} allocation), we define categories and use \textbf{predicate}(variable, starting-category, final-category). Categorical variables are represented as \textbf{toCategory}(variable). Quartiles are not used for \ac{prb} allocation as categories offer adjustable intervals for operator goals. Here, we define categories such that each category covers 10\% of the available \ac{prb} range. Mathematically, the interval for category \( C_i \) is defined as \(\left[ \frac{(i-1) \times \text{PRB}}{10}, \frac{i \times \text{PRB}}{10} \right)\), where \( i \) ranges from 1 to 10 and \ac{prb} is 50. For space reasons, we only present results for TRF1 in \S\ref{sec:evaluation}.\vspace*{4pt}

    \item \label{agent-rice} \textbf{Massive MIMO Scheduling Agent} \cite{MIMO-Scheduler-agent}: 
    This agent focuses on resource scheduling in Massive MIMO networks to maximize spectral efficiency while maintaining fairness among users. We evaluate two \ac{drl} agents: the SMART scheduler based on~\ac{sac}~\cite{haarnoja2018soft} proposed by the authors~\cite{MIMO-Scheduler-agent}, and a \ac{dqn}-based implementation~\cite{mnih2013playing}. The evaluation uses the Indoor Mobility Channel Measurements dataset~\cite{MIMO-Scheduler-agent}, providing channel state information for \ac{los} and \ac{nlos} scenarios, as well as low-speed and high-speed mobility profiles. The agent's state space includes \acp{kpi} such as \ac{mse}, \ac{dtu}, and \ac{g}. User's~\ac{g} are assigned based on their channel state correlation.

    For \ac{fol}-based symbolic representation, we focus on groups rather than individual users to assess if the agent learns to avoid interference from joint scheduling of different group's users. Continuous variables (\ac{mse}, \ac{dtu}) use \textbf{predicate}(variable, Quartile). The agent's decision is represented as \textbf{sched}(group\_number, quartile, percentage), where percentage is the proportion of scheduled users in the group and quartile is calculated for the number of scheduled users in the group with respect to the total number of users. For \ac{dtu} and user scheduling, we also use the MAX quartile to represent the highest observed value up to the current timestep, provided by the P2 algorithm~\cite{p2-algo}. This approach enables comparison of decisions across groups and assesses interference mitigation.

\end{enumerate}

The diverse nature and purpose of these agents allow us to assess \toolname's capabilities across different network management scenarios and \ac{drl} architectures.

Table~\ref{tab:action-space-size-comp-our-vs-explora} demonstrates the efficiency of \toolname in converting the agent's action space to symbolic representations, compared to \explora~\cite{explora}, which operates on low-level numerical representations. The \ac{fol} definition of \toolname requires approximately 99.5\% fewer nodes to model the behavior of Agent \ref{agent-neu} and 40\% fewer nodes for Agent \ref{agent-rice}. Unlike \explora, where the node count increases as the agent discovers new states and actions, \toolname's~\ac{kg} is bounded by the \ac{fol} definition. This enables \toolname to generate bounded~\ac{kg}s, regardless of action space size.

\begin{table}[bp]
    \centering
    \small
    \caption{Comparing Action Space Sizes}%
    \label{tab:action-space-size-comp-our-vs-explora}%
    \vspace*{-1ex}%
    \begin{tabularx}{\linewidth}{@{}>{\centering\arraybackslash\hsize=0.3\hsize}X>{\centering\arraybackslash\hsize=1.5\hsize}X>{\centering\arraybackslash\hsize=1.2\hsize}X@{}}
        \toprule
        \textbf{Agents} & \textbf{\toolname} & \textbf{\explora} \\
        \midrule
        A1 & \footnotesize $|\mathcal{L}| \cdot |\text{Pred}_{\text{PRB}}| \cdot |\text{Cat}_{\text{PRB}}| \cdot |\text{toPolicy}|$ & \footnotesize $\binom{\text{PRB}-1}{2} \cdot |\text{SP}|^{|\mathcal{L}|}$ \\[1ex]
        & \textit{Node Count} = 180
        & \textit{Node Count} = 31,752
        \\
        \midrule
        A2 & \footnotesize $|\mathcal{G}| \cdot |\mathcal{Q}| \cdot |\text{Percentage}|$ & \footnotesize $2^N$ \\[1ex]
        & \textit{Node Count} = 75
        & \textit{Node Count} = 128
        \\
        \bottomrule
    \end{tabularx}
    \vspace*{-2ex}%
\end{table}

\subsection{Understanding Agents' Behavior}
\label{subsec:insights}
We analyze the explanations generated by \toolname for agents \ref{agent-neu} and \ref{agent-rice}.
\begin{figure*}
\centering
\subfloat[Slice \ac{embb}\label{fig:neu-prob-dist-trf1-embb}]{%
	\includegraphics[width=.85\textwidth,keepaspectratio]{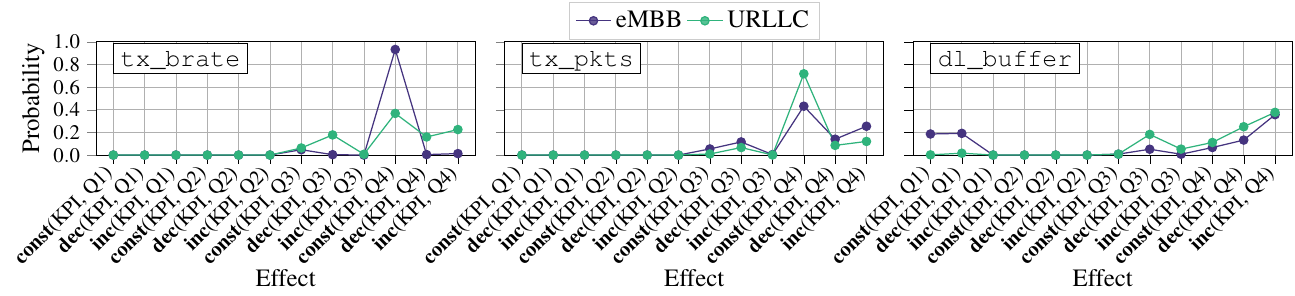}}%
\par
\vspace*{-2ex}%
\subfloat[Slice \ac{urllc}\label{fig:neu-prob-dist-trf1-urllc}]{%
	\includegraphics[width=.85\textwidth,keepaspectratio]{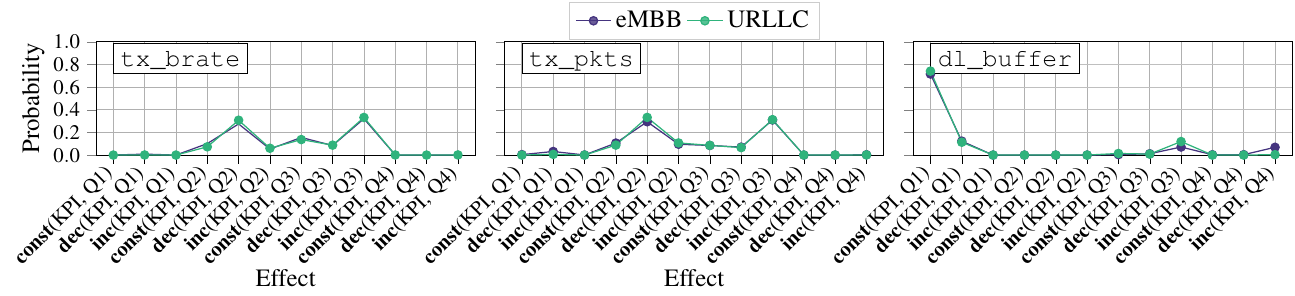}}%
\vspace*{-1ex}%
\caption{Probabilistic analysis of \toolname showing the probability distribution of decision effects for agent A1's variants under TRF1.}
\label{fig:prob-distr}%
\vspace*{-3ex}%
\end{figure*}

\begin{figure*}
    \centering%
    \includegraphics[width=0.95\textwidth, keepaspectratio]{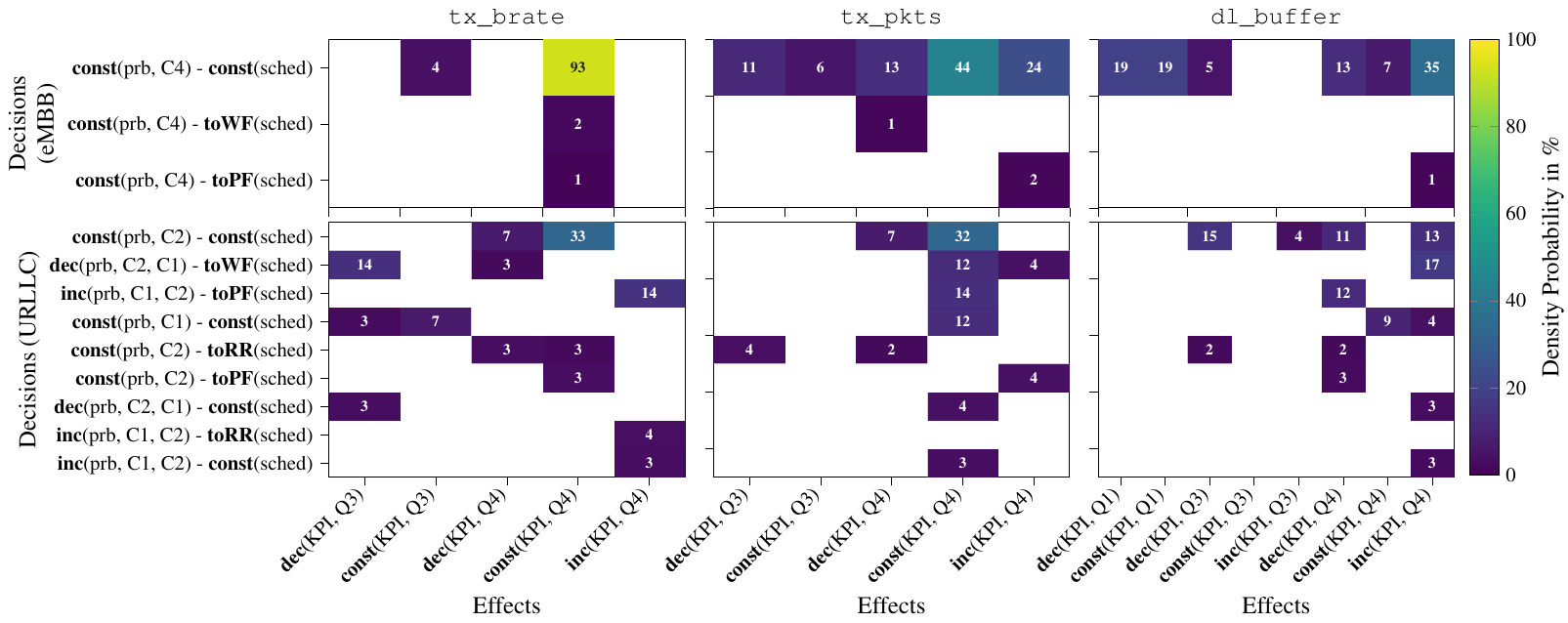}%
    \vspace*{-2ex}%
    \caption{Correlation density map of input \acp{kpi} and decisions for two variants of agent \ref{agent-neu} favoring \ac{embb} (first row) or \ac{urllc} (second row) slices.}
    \vspace*{-3ex}%
    \label{fig:neu-agent-heatmap-comp}%
\end{figure*}

\vspace*{4pt}
For agent \ref{agent-neu}, Fig.~\ref{fig:prob-distr}\subref{fig:neu-prob-dist-trf1-embb}~and~\ref{fig:prob-distr}\subref{fig:neu-prob-dist-trf1-urllc} illustrate the probability distribution of symbolic effects (changes in input \acp{kpi} due to the agent's decisions) for both \ac{embb} and \ac{urllc} slices under TRF1 traffic. The figures provide insights into how each variant of the agent manages different slices under the same traffic profile. Key observations are:

\begin{enumerate}[label=\textit{O\arabic*:}, wide=0\parindent, listparindent=0pt, align=left]
    \item \textit{High Throughput Maintenance for \ac{embb} Slice:} The \ac{embb} agent effectively maintains high throughput (\txbrate) for the \ac{embb} slice, as shown by the high probability of \txbrate remaining constant in Q4 ($\text{const}(\text{KPI}, \text{Q4})$) and low probability of other changes.

    \item \textit{Buffer Management in \ac{urllc} Slice:} Both agents aim to keep buffer occupancy (\dlbuff) low in the \ac{urllc} slice, with the \ac{urllc} agent performing slightly better, indicated by a higher probability of $\text{const}(\text{KPI}, \text{Q1})$ for this agent versus the \ac{embb} agent's counterpart.

    \item \textit{Takeaway:} The performance of the \ac{urllc} agent is not significantly superior to the \ac{embb} agent for the \ac{urllc} slice, suggesting that merely adjusting the reward function weights is insufficient for optimal agent action tuning.
\end{enumerate}


To further analyze the behavior of the two variants of agent \ref{agent-neu}, we examine the correlation density map of its decisions and their effects on the \ac{embb} slice (Fig.~\ref{fig:neu-agent-heatmap-comp}). In \ac{embb} mode, the agent consistently allocates high \ac{prb} resources ($\text{const}(\text{\ac{prb}}, \text{C4})$, 93\% density) and adapts to traffic variations mainly through scheduling policy adjustments. Conversely, in \ac{urllc} mode, the agent exhibits more diverse decision frequencies across \ac{prb} allocations and scheduling policies, indicating a less stable approach than the \ac{embb} counterpart.

\begin{figure}
\centering
\subfloat[Strategy: DQN\label{fig:rice-ag-dqn}]{%
\includegraphics[width=.99\columnwidth,keepaspectratio]{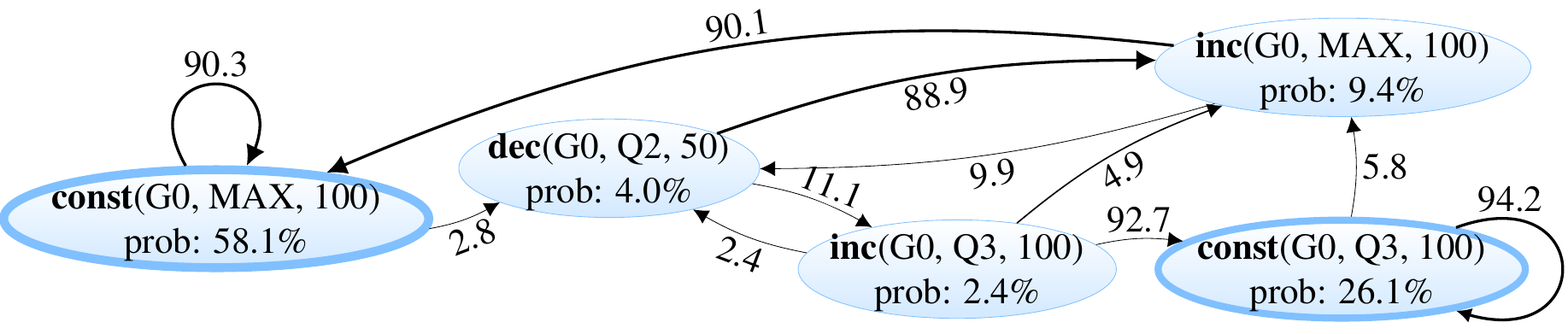}}%
\par
\vspace*{-1ex}%
\subfloat[Strategy: SAC\label{fig:rice-ag-sac}]{%
\includegraphics[width=.99\columnwidth,keepaspectratio]{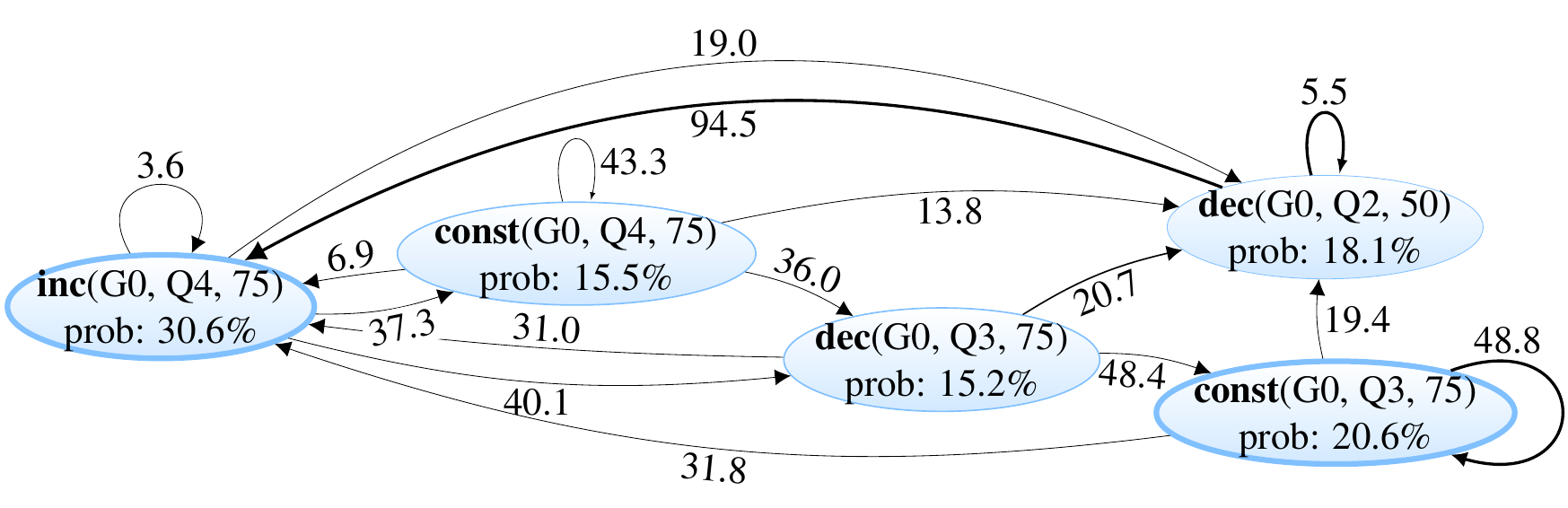}}%
\vspace*{-0ex}%
\caption{\ac{kg}s produced by \toolname for each implementation of agent \ref{agent-rice}.}%
\label{fig:example-graphs-rice}%
\vspace*{-3ex}%
\end{figure}

Let us now move to agent \ref{agent-rice}. Fig.~\ref{fig:example-graphs-rice} compares the \ac{kg}s for \ac{sac} and \ac{dqn} models in Group 0, revealing their learned strategies. The \ac{sac} agent (Fig.~\ref{fig:example-graphs-rice}\subref{fig:rice-ag-sac}) shows a stochastic decision-making pattern by distributing actions across various quartiles. One frequent transition path (Q2 $\rightarrow$ Q4 $\rightarrow$ Q3) indicates fluctuations between scheduling fewer and more users. In contrast, the \ac{dqn} agent (Fig.~\ref{fig:example-graphs-rice}\subref{fig:rice-ag-dqn}) demonstrates a more deterministic approach, with a dominant node representing a 58\% probability of scheduling users in the MAX quartile and a 90\% likelihood of maintaining this action. This suggests the \ac{dqn} agent's deterministic policy is more efficient for Group 0 scheduling, consistently maximizing the number of scheduled users. \toolname's \ac{kg}s provide valuable insights into agent decision trends, highlighting the effectiveness of deterministic policies in this scenario.

\begin{figure}
\centering
\subfloat[Mean DTU\label{fig:dtu-prob-distr}]{%
\includegraphics[width=.95\columnwidth,keepaspectratio]{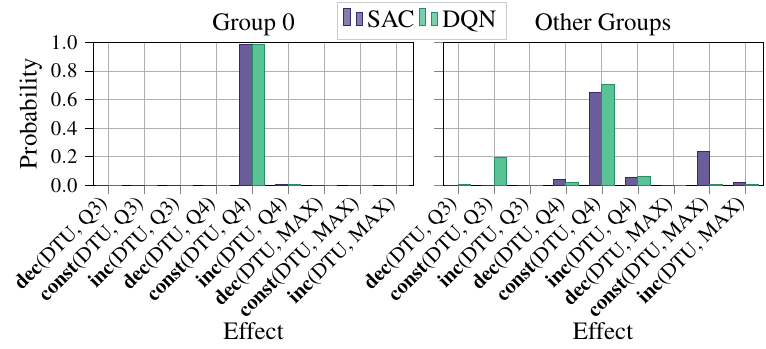}}%
\par
\vspace*{-1ex}%
\subfloat[Mean MSE\label{fig:dtu-mse-distr}]{%
\includegraphics[width=.95\columnwidth,keepaspectratio]{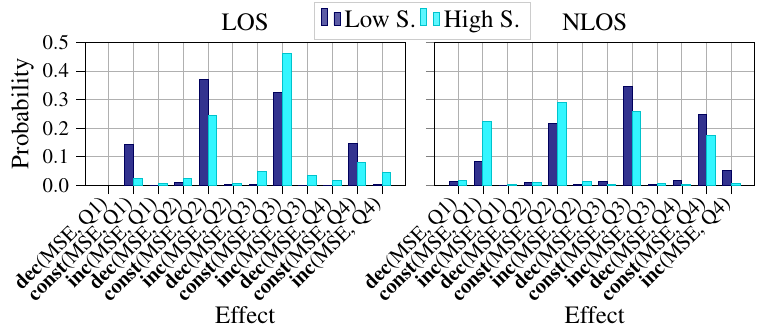}}%
\vspace*{-1ex}%
\caption{Agent \ref{agent-rice}: Probability distribution of input KPIs (\ac{dtu} and \ac{mse})}
\label{fig:analysis-agent-rice}%
\vspace*{-2.5ex}%
\end{figure}

Fig.~\ref{fig:analysis-agent-rice} shows the symbolic effect analysis of agent \ref{agent-rice} using \toolname, focusing on \ac{dtu} and \ac{mse}. Fig.~\ref{fig:analysis-agent-rice}\subref{fig:dtu-prob-distr} depicts the probability distribution of mean \ac{dtu} for \ac{sac} and \ac{dqn} models. For Group 0, both agents consistently achieve mean \ac{dtu} in Q4, indicating efficient scheduling. For other groups, the distribution varies: \ac{sac} favors Q4 or MAX quartiles, while \ac{dqn} leans towards Q3 or Q4. Similarly, Fig.~\ref{fig:analysis-agent-rice}\subref{fig:dtu-mse-distr} presents the mean \ac{mse} distribution across different channel conditions (\ac{los} and \ac{nlos}) and mobility profiles (low and high speed). In \ac{los} scenarios, \ac{mse} is concentrated in Q2 and Q3 for both speed profiles. \ac{nlos} conditions show a broader distribution. High-speed scenarios exhibit mean \ac{mse} spread across all quartiles, indicating higher variability, while low-speed scenarios concentrate in Q3 and Q4. This analysis reveals the impact of user speed on \ac{mse} distribution, especially in \ac{nlos} conditions. While this may seem obvious to experts, \toolname is crucial for explaining how these effects unfold within the \ac{drl} agent's decision process.

\subsection{Improving Agent's Behavior}
\label{subsec:action-steering}

This section analyzes the performance of \ac{ias} (module~\ding{184} in Fig.~\ref{fig:xrlogicLens-architecture}), focusing on agent \ref{agent-rice}. We present quantitative analyses for three key use cases described in \S\ref{subsec:intent-action-steering} to demonstrate \toolname's capability to enhance agent performance in complex network scenarios.
 
\subsubsection{Reward Maximization}\label{subsub:reward-maximization}

\begin{figure*}
    \centering%
    \includegraphics[width=.99\textwidth,keepaspectratio]{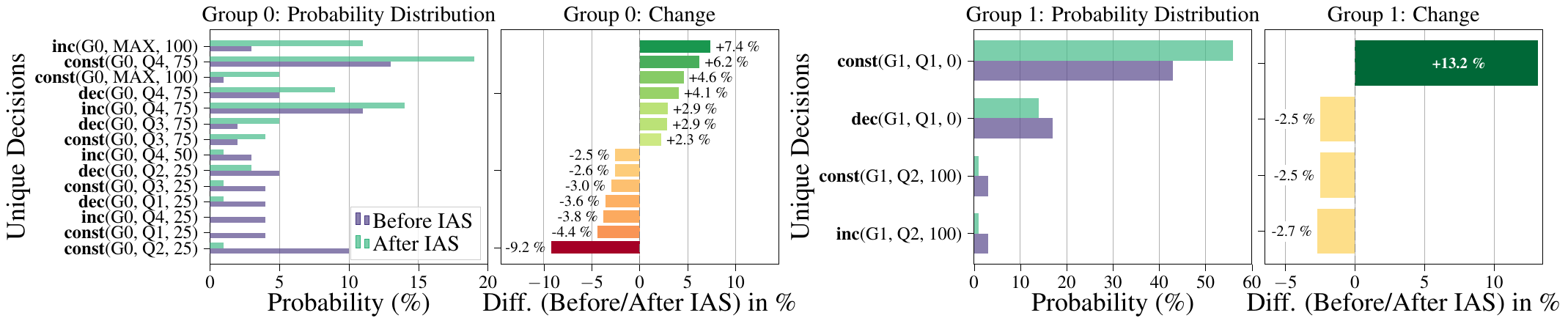}%
    \vspace*{-1ex}%
    \caption{Comparison of agent \ref{agent-rice} decisions before and after applying Action Steering \ac{ias}, showing the effect of \ac{ias} on agent behavior.}
    \label{fig:compare-before-after-action-steering-prob-dist}%
    \vspace*{-3ex}%
\end{figure*}

As discussed in Section~\ref{subsec:intent-action-steering}, \toolname \ac{ias} can automatically improve agent behavior to maximize cumulative reward~\eqref{eq:reward-maximizing-formula}. We apply reward maximization \ac{ias} to agent \ref{agent-rice} using the \ac{nlos} dataset with high-speed mobility over 10 test episodes. Fig.~\ref{fig:compare-before-after-action-steering-prob-dist} illustrates how this functionality affects the probability distribution of symbolic actions for the target agent. We observe that:

\begin{enumerate}[label=$\bullet$]
    \item \textit{Group 0}: Before \ac{ias}, the agent predominantly scheduled either 25\% or 75\% of users. After \ac{ias}, there is a notable increase in scheduling 75\% or 100\% of users, with a substantial reduction in scheduling 25\% of users.
    \item \textit{Group 1}: After \ac{ias}, the agent shows a 13.2\% increase in not scheduling any users from this group, indicating a strategy to avoid inter-group interference.
\end{enumerate}

Analysis of changes before and after applying \ac{ias} suggests that \ac{ias}'s reward maximization mode alters agent behavior towards more efficient resource use, reducing inter-group interference while maximizing utilization for Group 0.

\begin{figure}
    \centering%
    \includegraphics[width=.95\columnwidth,keepaspectratio]{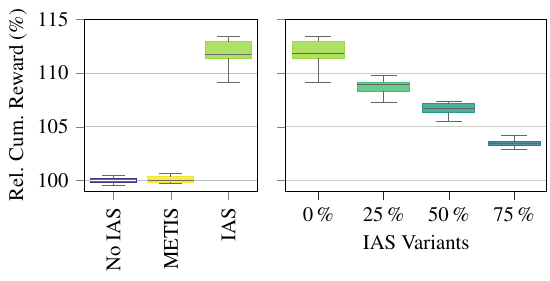}%
    \vspace*{-1.75ex}%
    \caption{Comparing relative cumulative reward of different benchmarks for Agent \ref{agent-rice}. "No \ac{ias}" indicates the basic agent, METIS shows improvement with \cite{metis-meng2020interpreting}, and the other boxplots show improvements with \toolname by starting \ac{ias} at different fractions of the test set (0\%, 25\%, 50\%, and 75\%).}
    \label{fig:sacg-cum-reward-comp-no-met-as}%
    \vspace*{-2ex}%
\end{figure}

Fig.~\ref{fig:sacg-cum-reward-comp-no-met-as} shows how these changes translate to cumulative reward~\eqref{eq:reward-maximizing-formula} improvement. Here, we benchmark \toolname against METIS~\cite{metis-meng2020interpreting}, which uses a decision tree-based approach to improve \ac{rl} agent performance. The figure reports the cumulative reward attained by the baseline \ref{agent-rice} agent, by the same \ref{agent-rice} agent improved by Metis, and by agent \ref{agent-rice} improved by \toolname's \ac{ias}.

We observe that \toolname achieves a median 11.76\% improvement in cumulative reward compared to the basic agent, whereas Metis only provides a 0.07\% gain. This highlights \toolname's ability to leverage symbolic knowledge effectively to identifying suboptimal decisions and replacing them with better alternatives from similar past states. Metis's marginal improvement suggests limitations in fully characterizing and exploiting agent knowledge.

The right portion of Fig.~\ref{fig:sacg-cum-reward-comp-no-met-as} shows the impact of initiating \ac{ias} at different points during the test episode. Starting late yields diminishing performance gains compared to starting early. While a late start can leverage more knowledge, it can only enact changes for a limited time, ultimately determining the lower cumulative reward. However, a late start yields reduced variance in performance improvements, due to more refined action replacements resulting from more knowledge gathered. These results show \toolname's significant potential to enhance \ac{rl} agent performance through knowledge-based action steering, outperforming other existing explainers and being effective even after a minimal amount of data has been collected about the target agent's behavior.

\subsubsection{Decision Conditioning}

\begin{figure}
\centering
\subfloat[Decision Conditioning: $\textbf{notSchedule}(a_t, \text{6})$\label{fig:ue-sched}]{%
	\includegraphics[width=.7\columnwidth,keepaspectratio]{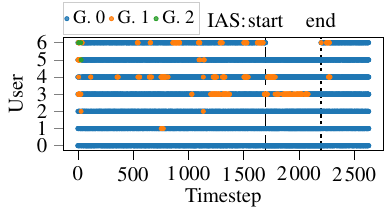}}%
\par
\vspace*{-2ex}%
\subfloat[Reward improvement with \ac{ias} versus direct forcing\label{fig:rew-sched-ue}]{%
	\includegraphics[width=.75\columnwidth,keepaspectratio]{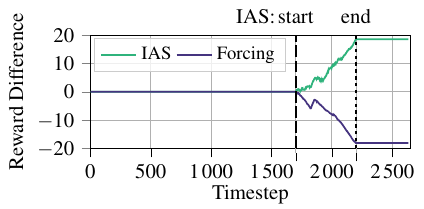}}%
\vspace*{-0ex}%
\caption{Intent-based operation of enforcing not scheduling a specific user with \ref{agent-rice}. \ac{ias} allows the agent to achieve high reward compared to blindly forcing actions.}
\label{fig:dec-cond}%
\vspace*{-3ex}%
\end{figure}

\ac{ias} can enforce high-level intents for flexible control of agent behavior without excluding reward maximization, as discussed in Section \ref{subsec:intent-action-steering}. This capability is crucial for network management scenarios with specific operational requirements like prioritization. We use \toolname to prohibit agent \ref{agent-rice} from scheduling user 6 for about 500 timesteps ($t \in [1700, 2200]$), achieved by conditioning the agent's behavior using the \ac{fol} term $\textbf{notSchedule}(a_t, \text{6})$. To apply this constraint, \ac{ias} uses both \ac{kg} and DB to find previously seen states similar to the current state where the agent's decision did not include scheduling user 6. If multiple timesteps satisfy the condition, we choose the action that provided the best reward when applied previously.

The outcome is shown in Fig.~\ref{fig:dec-cond}\subref{fig:ue-sched}, where user 6 is not scheduled in any group (denoted by different colors) in the target interval. Fig.~\ref{fig:dec-cond}\subref{fig:rew-sched-ue} compares the performance of \ac{ias} and direct action forcing i.e., explicitly removing user 6 from the agent's decision without considering any consequences of this change.

The cumulative reward differences attained by each technique with respect to the original agent \ref{agent-rice} show that:

\begin{enumerate}[label=$\bullet$] 
    \item \ac{ias} enhances the reward over time while applying constraints which indicates that~\ac{ias} can choose actions with the highest reward from the~\ac{kg} for replacement.

    \item Direct forcing leads to a rapid decline in cumulative reward compared to the baseline, indicating performance degradation. This is because removing one of the scheduled users reduces the spectral efficiency factor of the reward function.
\end{enumerate}

These results demonstrate \toolname's superiority in applying constraint. By leveraging symbolic knowledge, \ac{ias} adjusts the agent's actions to accommodate constraints while optimizing overall performance. This shows \toolname's potential to enhance the flexibility and efficiency of \ac{rl} agents in real-world telecommunication systems, balancing operational requirements with system optimization.

\subsubsection{Accelerated Learning}

\begin{figure}
    \centering%
    \includegraphics[width=.95\columnwidth,keepaspectratio]{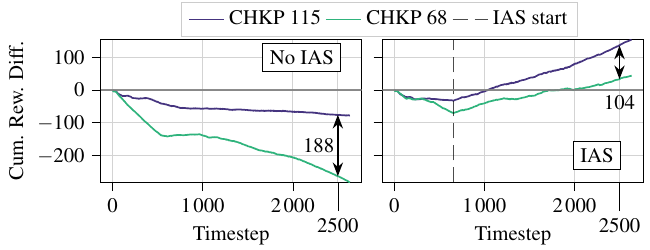}%
    \vspace*{-2ex}%
    \caption{Relative cumulative reward of CHKPs 115 and 68 compared to the final model (CHKP 205) without IAS (left) and with IAS enabled after the first 25\% of the test episode (right).}
    \label{fig:acc-learning-checkpoint-comp}%
    \vspace*{-3ex}%
\end{figure}

Fig.~\ref{fig:acc-learning-checkpoint-comp} demonstrates \ac{ias}'s ability to reduce training time for \ac{rl} agents. We compare agent \ref{agent-rice}'s performance at three training \acp{chkp}: the final model, used until now in the evaluation, is trained for 205 episodes (\ac{chkp} 205, the baseline). Next, we create variants with fewer training episodes, stopping the training earlier at episodes 68 and 115. Note that all these \acp{chkp} are in the stable region of cumulative reward upon training.

The left portion of Fig.~\ref{fig:acc-learning-checkpoint-comp} shows the cumulative reward difference of \ac{chkp} 68 and \ac{chkp} 115 relative to the baseline \ac{chkp} 205 without \ac{ias}. As expected, the cumulative reward without \ac{ias} is lower for \ac{chkp} 68 and \ac{chkp} 115 compared to \ac{chkp} 205. The right portion of the figure illustrates the impact of applying \ac{ias} after the first 25\% of the testing episode (denoted by the dashed line in the figure):

\begin{enumerate}[label=$\bullet$] 
    \item Before \ac{ias} for accelerated learning is enabled, the agents trained up to \acp{chkp} 68 and 115 yield a lower cumulative reward compared to the baseline, similar to the left plot.
    \item Upon \ac{ias} activation, the agents trained up to \acp{chkp} 68 and 115 start leveraging the \ac{kg} through \ac{ias} to replace low-reward actions with higher-reward ones, as detailed in Section~\ref{subsub:reward-maximization}. This effectively boosts cumulative reward with respect to the baseline agent and eventually surpasses it by the end of the episode.
    \item Using \ac{ias} by the end of the test episode reduces the performance gap between \ac{chkp} 115 and 68 by 45\% (from 188 to 104), indicating more consistent performance improvement across different training stages.
\end{enumerate}

These results show \toolname's ability to leverage an agent's acquired knowledge during its operation effectively, even at earlier training stages. \ac{ias}'s accelerated learning makes it possible for an agent trained up to \ac{chkp} 68 to perform as well as an agent trained up to \ac{chkp} 205, with a 66.7\% training time reduction, equivalent to 13 hours and 41 minutes. Both are trained on an NVIDIA A100 40 GB GPU cluster.

This suggests that \ac{drl} agents accumulate valuable knowledge early in training but may not fully exploit it without further assistance until later episodes. The \ac{ias}'s \textit{accelerated learning} strategy provides a mechanism to fully exploit this latent knowledge, thereby providing high performance.

\subsubsection{Summary of Insights}
The analyses of \textit{reward maximization}, \textit{decision conditioning}, and \textit{accelerated learning} collectively demonstrate the flexibility and effectiveness of \toolname's \ac{ias} approach. By leveraging symbolic knowledge, \ac{ias} not only improves overall agent performance but also enables precise control over agent behavior with flexible intents and makes early stopping of the training process viable. These capabilities are particularly valuable in dynamic network environments, where adaptive management and efficient resource utilization are crucial.

\section{Related Work}
\label{sec:relworks}
Relevant to our work are studies on the use of \ac{drl} for mobile networking, \ac{xai} at large, and \ac{xrl} techniques, especially those applicable to mobile networking.

\simpletitle{\ac{drl} in Mobile Networking} \simpletitle{\ac{drl} in Mobile Networking} \ac{drl} models have gained prominence for handling high-dimensional data in dynamic environments, excelling at rapid parameter reconfiguration during exploitation~\cite{ge-chroma-mobicom23}. Unlike \acp{dt} which suit rule-based configurations using historical data (e.g., Auric for \acp{bs}~\cite{auric}, Configanator for content-delivery~\cite{naseer2022configanator}), \ac{drl} applications span diverse areas: dynamic spectrum access~\cite{naparstek2018deep}, joint user scheduling with antenna allocation~\cite{naeem2022optimal} and mmWave configuration~\cite{zhang2022reinforcement}, resource management~\cite{vrain-andres-mobicom19}, network slicing~\cite{twc-23-drl-resourcealloc,polese2022colo}, mobility management~\cite{ho2022joint,ma2020aesmote}, service coverage~\cite{yang2018decco}, and anomaly detection~\cite{ma2020aesmote,CAMINERO201996}.

\simpletitle{XAI For Mobile Networks} Future 6G networks embrace the vision for native, explainable network intelligence. Seminal works~\cite{xai-6g,li-chen-xai-6g} motivate the need for \ac{xai} and stress that the lack of explainability may lead to poor AI/ML model design. This has been proved detrimental in the presence of adversarial attacks~\cite{deexp23}. All areas where \ac{ai} is applied to mobile networking tasks can benefit from explainability. These include physical and MAC layer design, network security, mobility management, and localization~\cite{challita-xai-examples}.

\simpletitle{Related Works on \ac{xrl}} The analysis of transitions between actions is not new~\cite{pmlr-v119-gottesman20a,amir2018highlights}. HIGHLIGHTS~\cite{amir2018highlights} produces summaries of agent behavior for general audience, while~\cite{pmlr-v119-gottesman20a} focuses on expert explanations by highlighting influential transitions whose removal significantly affects rewards. However, this approach fails with autoencoders masking real input as in Agent~\ref{agent-neu}. DeepSynth~\cite{deepsynth} reveals patterns in reward sequences to reduce exploration time. PIRL~\cite{pmlr-v80-verma18a} generates interpretable and verifiable policies using domain-specific languages, but designing primitives per \ac{ran} scenario is inefficient unlike \toolname's symbolic representation. The work~\cite{alden-llms-xrl} uses reward decomposition with \ac{llm} for text explanations. EXPLORA~\cite{explora} is closest to our work but synthesizes network-aware explanations using attributed graphs rather than symbolic AI like \toolname.

\section{Conclusions}%
\label{sec:conc}%

In this paper, we proposed \toolname, a new \ac{xrl} technique that generates explanations for \ac{drl} agents. \toolname leverages symbolic AI to represent concepts and their relationships, coupled with logical reasoning. In this way, \toolname provides a competitive advantage over existing explainers because it clarifies how \ac{drl} agents arrive at their decisions in an understandable manner to the human observer.

We validated our approach extensively with agents trained on both real-world datasets and datasets generated through Colosseum. We demonstrated that \toolname not only improves the clarity of the explanations but also enables performance improvements for the agents by using the knowledge generated by the explanations. Specifically, we show that intent-based action steering \ac{ias} achieves a median 12\% improvement in cumulative reward over the baseline \ac{drl} solution. The authors have provided public access to their code and/or data at https://github.com/RAINet-Lab/symbxrl

\section*{Acknowledgment}
This work is partially supported by bRAIN project PID2021-128250NB-I00 funded by MCIN/ AEI /10.13039/501100011033/ and the European Union ERDF ``A way of making Europe''; by Spanish Ministry of Economic Affairs and Digital Transformation, European Union NextGeneration-EU/PRTR projects MAP-6G TSI-063000-2021-63, RISC-6G TSI-063000-2021-59 and AEON-ZERO TSI-063000-2021-52; C. Fiandrino is a Ramón y Cajal awardee (RYC2022-036375-I), funded by MCIU/AEI/10.13039/501100011033 and the ESF+.

\ifCLASSOPTIONcaptionsoff
\newpage
\fi


\bibliographystyle{IEEEtran}
\bibliography{biblio}

\end{document}